\documentclass[preprint]{aastex}
\usepackage{emulateapj5}

\usepackage{graphicx}
\usepackage{amsmath}
\newcommand{\flux}{{\rm\ erg\ cm^{-2}\ sec^{-1}}}

\begin{document}

\title{Direct Detection of Warm Dark Matter in the X-ray} 
 
\author{Kevork Abazajian\altaffilmark{1}, George M. Fuller\altaffilmark{1}, and Wallace H. Tucker\altaffilmark{1,2}}
\altaffiltext1{Department of Physics and Center for Astrophysics and
Space Sciences, University of California, San
Diego, La Jolla, California, 92093-0319}
\altaffiltext2{Harvard-Smithsonian Center for Astrophysics, 60 Garden Street, Cambridge, MA 02138}

\begin{abstract}
We point out a serendipitous link between warm dark matter (WDM)
models for structure formation on the one hand and the high
sensitivity energy range (1-10 keV) for x-ray photon detection on the
{\it Chandra} and {\it XMM-Newton} observatories on the other.  This fortuitous
match may provide either a direct detection of the dark matter or
exclusion of many candidates.  We estimate expected x-ray fluxes from
field galaxies and clusters of galaxies if the dark matter halos of
these objects are composed of WDM candidate particles with rest masses
in the structure formation-preferred range ($\sim$1 keV to $\sim$20
keV) and with small radiative decay branches. Existing observations
lead us to conclude that for singlet neutrinos (possessing a very
small mixing with active neutrinos) to be a viable WDM candidate they
must have rest masses $\lesssim 5\,\rm keV$ in the zero lepton number
production mode.  Future deeper observations may detect or exclude the
entire parameter range for the zero lepton number case, perhaps
restricting the viability of singlet neutrino WDM models to those
where singlet production is driven by a significant lepton number.
The Constellation X project has the capability to detect/exclude
singlet neutrino WDM for lepton number values up to 10\% of the photon
number. We also consider diffuse x-ray background constraints on these
scenarios.  These same x-ray observations additionally may constrain
parameters of active neutrino and gravitino WDM candidates.
\end{abstract}


\section{Introduction}

In this paper we show how the {\it Chandra}, {\it XMM-Newton}, and future
Constellation X observatories can detect or exclude several Warm Dark
Matter (WDM) candidates, including singlet (``sterile'') neutrinos,
heavy active neutrinos, and gravitinos in some models. In essence, we
show here how the technology of modern x-ray astronomy allows the
exploration of a new sector of particle physics, one where interaction
strengths could be characteristically some 10 orders of magnitude
weaker than the Weak Interaction.  Processes with these interaction
strengths likely never could be probed directly in a laboratory.  

The x-ray observatories, however, possess several advantages when it
comes to probing WDM candidate particles.  First, the sensitive energy
range for x-ray photon detection on these instruments is $\sim$1\ keV
to $\sim$10\ keV.  Serendipitously, this is more or less coincident
with the WDM candidate particle rest mass range which is preferred in
studies of structure formation \citep*{Bode2000}.  Some models posit
the production of WDM particles through their tiny intractions with
ordinary matter.  In several cases these very weak interactions lead
to small radiative decay branches, producing photons with energies of
order the WDM particle rest mass.

This is where the second advantage of x-ray astronomy comes in.  Even
though the WDM particle decay rate into photons can be very small
(lifetimes against radiative decay are typically $\sim\!
10^{16}\times$ Hubble time), the dark matter halos of galaxies and
galaxy clusters can contain huge numbers of particles, e.g., in
the latter class of objects some $10^{79}$ particles of rest mass
$\sim$1 keV.  In effect, dark matter halos can serve as laboratories
of enormous ``fiducial volumes'' of dark matter particles.

The first evidence for dark matter was the velocity dispersion of
galaxies in the Coma Cluster, which required mass to light ratios in
the cluster to exceed those inferred for our Galaxy by many times
\citep{Zwicky1933}.  Later, observations of giant spiral galaxies implied
that their disks are imbedded in larger halos of dark matter
\citep*{Ostriker1974,Einasto1974}.  Recently, problems in cosmological
structure formation models have led to interest in alternatives to
the standard Cold Dark Matter (CDM) model for structure formation.

The primary problem encountered in comparing calculations of structure
formation in CDM models to observation is that simulations predict a
large overabundance of small halos near galaxies such as our own.
Structure formation in these models occurs through hierarchical growth
of fragments into larger objects.  This hierarchical structure is
obvious in clusters of galaxies, where the numerous constituent
galaxies are seen directly.  Individual galaxy formation is also
hierarchical in CDM simulations.  These simulations predict a large
number of dark matter subhalos, about 500, for each Milky-Way type
halo \citep{Moore:1999wf,Ghigna1999}; however, only 11 dwarf galaxies
are observed near our Galaxy.  The observed paucity of such
substructure (dwarf galaxies) has been interpreted as a fundamental
failure of CDM models.  The hallmark of {\it Cold} Dark Matter
particles is a very small collisionless damping (free streaming)
scale, i.e., considerably smaller than the scale associated with
dwarf galaxies, $\sim\! 0.3{\rm\ Mpc}\ (\sim\! 10^{10}\ M_\odot)$.
Between the large free streaming scale ``top-down'' Hot Dark Matter
(HDM) structure formation scenarios and the ``bottom-up'' scenarios of
CDM models, lies the intermediate regime of WDM
\citep*{Colombi1996,Bode2000}, with typcial dark matter particles of
rest mass $\sim\!  1 \rm\, keV$.  

On the other hand, the lack of dwarf galaxies may be caused by
feedback processes from supernovae and heating in small halos because
of the initial formation of very massive stars in a zero-metallicity
environment \citep{Bullock2001,Binney2001,Abel1998}.  However, it is
not clear that these processes would be successful in disrupting dwarf
galaxy formation. Nevertheless, there remain many mysteries regarding
the earliest stars.  Observations of abundances in ultra-metal-poor
halo stars may be giving us some new insights into these issues.  For
example \citet{Qian2001} have argued that an inferred increase in $\rm
[Fe/H]$ with no concomitant increase in $r$-process abundances may
signal the activity of very massive objects.

Another potential problem in CDM models is that the universal density
profiles predicted in simulations of structure formation have a
monotonic increase of density towards the center of halos.  This could
give a central ``cusp,'' that is, a discontinuity in the derivative of
the spatial density distribution that is a result of the initial
singularity in velocity dispersion.  The observational searches for a
centrally-peaked dark matter profile are as yet inconclusive
\citep{Swaters2000,vandenbosch2000a}. However, the measurement of the
innermost rotation curve in dark-matter-dominated galaxies may provide
the dark matter profile of these halos.

One proposed solution to the central density problem is
self-interacting dark matter \citep{Spergel:1999mh}, which has Strong
Interaction strength forces among dark matter particles and
essentially no interactions between dark matter particles and ordinary
matter.  This kind of interaction would soften cores as a result of
efficient energy exchange in halo centers.

In fact, however, \citet{Dalcanton:2000hn} find that the core phase
space density distributions in dwarf galaxies and clusters may behave
as simple power laws over eight orders of magnitude in phase
density. They argue that such density profiles could not arise from
either a WDM or self-interacting dark matter scenario [see also
\citet{Sellwood2000}].

The idea of using discrete UV photon sources to place limits on the
radiative decay of active neutrino HDM was first proposed by
\citet{Shipman1981}.  The lack of UV photons from the rich cluster
A665 was used by \citet{Melott1994} to constrain a decaying neutrino
dark matter model proposed by \citet{Sciama1990}.  Our approach to the
detection of a radiative flux from cluster cores is similar: we
explore the efficacy of x-ray observations in obtaining 
detections of or constraints on singlet neutrino, active
neutrino, and gravitino dark matter candidates.

If observations fail to find the decay flux predicted for a specific
dark matter candidate, then the decay constraints presented here can
provide upper mass bounds on dark matter candidates. Together with
existing structure-derived lower limits on the dark matter particle
rest mass in
these models, we can potentially exclude specific particle dark matter
candidates.  

For example, from observation of the power spectrum of the
Lyman-$\alpha$ forest clouds at high redshift it can be concluded that
there is significant structure on small scales.  This requires a small
collisionless damping scale associated with a dark matter particle
with a thermal energy spectrum, and in this limit the particle's rest
mass must be greater than 750 eV for a standard warm dark matter
particle which decoupled at high temperature \citep{Narayanan2000}.
In addition, a paucity of power on small scales can delay the
formation of structure at high redshifts and delay cosmological
reionization.  Such considerations corroborate the Lyman-$\alpha$
forest constraints and also favor a WDM particle to have a rest mass
greater than 750 eV \citep{Barkana2001}.  The energy distribution for
singlet neutrino relics is different from dark matter which has
decoupled at high temperature, and is generally ``warmer.''
\citet{Colombi1996} find that the power spectrum for a sterile
neutrino with rest mass $m_s$ is the same for a standard WDM relic
particle with rest mass $m_X$ if $m_s\approx 2.6 m_X$.  Therefore,
current lower limits on singlet WDM particle rest mass favor singlet
neutrinos with masses $m_s \gtrsim 2.0\rm\ keV$.

In Section 2, we outline the radiative decay rates that can be probed
in deep x-ray observations.  Section 3 describes the singlet neutrino,
active neutrino, and gravitino WDM candidates and their radiative
decay modes. In Section 4, we describe specific limits from field
galaxy dark matter halos and those from clusters of galaxies.  Section
5 presents current diffuse limits on singlet neutrino dark matter.  In
Section 6, we present our conclusions.

\section{Dark Matter Halos as Particle Reservoirs}

An object such as a field galaxy or dwarf galaxy or a cluster of
galaxies possessing a dark matter halo of mass $M_{\rm DM}$ will be
composed of $N=M_{\rm DM}/m_X$ dark matter particles of rest mass
$m_X$.  If $\Gamma_\gamma$ is the dark matter particle decay rate into
photons of energy $E_\gamma$, then the total associated x-ray
luminosity is 
\begin{equation}
{\cal L} \approx \frac{E_\gamma}{m_X} M_{\rm DM}\ \Gamma_\gamma.
\label{lumgen}
\end{equation}
Here we have assumed that the halo is relatively nearby and that
redshift effects on the luminosity are negligible.

For illustration let us take the case where $E_\gamma = m_X/2$ (here,
and unless mentioned otherwise, we adopt units where $\hbar=c=1$).
The flux from an object is simply $F= {\cal L}/4\pi D_L^2$, where
$D_L$ is the luminosity distance to the object.  With a reasonable
integration time observation of a dark matter halo, a line of energy
$E_\gamma=m_X/2$ can be detected at a flux above, for example, $F_{\rm
det}= 10^{-13}\ \flux$.  (We will show that this actually is the
appropriate limit for {\it Chandra}.)  This can place a limit on the
radiative decay rate of the relic dark matter particle at
\begin{eqnarray}
\Gamma_\gamma \lesssim && (2.4\times 10^{20}{\ \rm yr})^{-1} 
\left(\frac{F_{\rm det}}{10^{-13}\flux}\right)\cr
&&\times\left(\frac{M^{\rm fov}_{\rm DM}}{10^{11}\,M_\odot}\right)^{-1}
\left(\frac{D_L}{1\,\rm Mpc}\right)^{2},
\label{decaylimit}
\end{eqnarray}
where $M^{\rm fov}_{\rm DM}$ is the total mass of dark matter within
the observed field of view.  As we show below, a decay rate limit of
this magnitude can be significant in constraining singlet neutrino,
active neutrino, and gravitino WDM candidate parameters.

\section{Warm Dark Matter Particle Candidates}
\subsection{Singlet Neutrinos}

In general, the neutrino mass eigenstates $\nu_a\
(a=1,2...)$ are related by a unitary transformation to the flavor
eigenstates $\nu_\alpha\ (\alpha=e,\mu,\tau,s...)$
\begin{equation}
\nu_a = \sum_\alpha U_{a \alpha} \nu_\alpha.
\label{transformnu}
\end{equation}
A singlet or ``sterile'' neutrino, $\nu_s$, that has a very small
mixing, $\sin^2 2\theta \approx 4 |U_{1 s} U_{2 s}|^2 \ll 1$, with one
or more doublet (``active'') neutrinos, could be produced
non-thermally via active neutrino scattering in the early
universe. This was proposed as a WDM candidate by
\citet{Dodelson:1994je}.  Singlet neutrino dark matter also could be
produced by matter-enhancement [a Mikheyev-Smirnov-Wolfenstein
resonance \citep{Mikheyev:1985,Wolfenstein:1978ue}] driven by a
primordial net lepton number residing in the active neutrino seas
\citep{Shi:1998km}. Interestingly, the singlet neutrino could be
produced in the requisite numbers to be a WDM candidate in these
scenarios even for extremely small vacuum mixing angles, $10^{-13}
\lesssim \sin^2 2\theta \lesssim 10^{-7}$.

Supernova constraints on these scenarios were considered by George
M. Fuller.\footnote{Fuller, G.\ M.\ 2000, Neutrino Astrophysics and
Cosmology (lectures at the XXVIII SLAC Summer Institute on Particle
Physics: Neutrinos from the Lab, the Sun, and the Cosmos). Available
at http://www.slac.stanford.edu/gen/meeting/ssi/2000/fuller.html.}
\citet{Dolgov:2000ew} did another calculation of the
\citeauthor{Dodelson:1994je} non-resonant scattering production
scenario for WDM singlets and also discussed diffuse photon background
and SN 1987A limits on these models.

Singlets with rest mass $m_s \gtrsim 1\rm\ keV$ are produced during or
prior to the quark-hadron transition.  In this case, the effects of
dilution, enhanced scattering rates, and the evolution of the thermal
potential become important. \citet*{Abazajian2001} (AFP) considered
singlet WDM production in both the non-resonant
\citeauthor{Dodelson:1994je} and resonant \citeauthor{Shi:1998km}
modes and explicitly took account of these early universe
thermodynamic effects.  AFP also considered in detail constraints on
these models arising from diffuse photon backgrounds, cosmic microwave
background, $^6$Li and $^2$H, Big Bang Nucleosynthesis, and supernova
effects limits.

In these models the singlet neutrino is produced in the early universe
through non-equilibrium scattering processes involving active
neutrinos and other weakly interacting particles. (Singlet neutrinos
could be produced via coherent MSW resonance in one limit of
matter-endhancement scenarios.)  In the matter-enhanced, resonant
singlet neutrino production scenarios, the relic density and energy
spectrum of the singlet neutrinos can depend on the initial neutrino
lepton number residing in the active neutrino seas
\begin{equation}
L_{\nu_\alpha} \equiv \frac{n_{\nu_\alpha} -
n_{\bar\nu_\alpha}}{n_\gamma},
\end{equation}
where $n_{\nu_\alpha} (n_{\bar\nu_\alpha})$ is the neutrino
(antineutrino) number density and $n_\gamma= 2\zeta(3)T^3/\pi^2
\approx 0.243T^3$ is the photon number density. In the case of small
initial neutrino lepton number (small here means the same order of
magnitude or smaller than the baryon-to-photon ratio $\eta\sim
10^{-10}$), ${L}_{\nu_\alpha}\approx 0$, the production occurs at
larger mixing angles than in the non-standard, yet plausible case, of
large lepton number, $0.001 \lesssim L_{\nu_\alpha} \lesssim 1$.  For
large lepton numbers, production is resonantly enhanced at low
energies and the energy spectrum of the singlet neutrinos can be
``cooler'' (that is, possess a smaller collisionless damping scale)
than for the case of a thermal energy spectrum \citep{Shi:1998km}.

For the $L_{\nu_\alpha}\approx 0$ case, the fraction of the
closure density in singlet neutrinos that is produced in the early
universe was found by AFP to be approximately 
\begin{equation}
\Omega_{\nu_s} h^2 \approx 0.3\ 
\left(\frac{\sin^2  2\theta}{10^{-10}}\right)
\left(\frac{m_s}{100\rm\, keV}\right)^2.
\label{omega}
\end{equation}
Here $\theta$ is the vacuum mixing angle defined by an {\it
effective} two-by-two unitary transformation between active
$\nu_\alpha$ species and a singlet species $\nu_s$:
\begin{eqnarray}
|\nu_\alpha \rangle &=& \cos \theta |\nu_1\rangle + \sin \theta | \nu_2
 \rangle \cr
|\nu_s \rangle &=& -\sin \theta |\nu_1\rangle + \cos \theta | \nu_2
 \rangle,
\end{eqnarray}
where $ |\nu_1\rangle $ and $| \nu_2 \rangle$ represent neutrino
energy (mass) eigenstates corresponding to vacuum mass eigenvalues
$m_1$ and $m_2$, respectively.  Here we define $h=H_0/(100\rm\ km\
s^{-1}\ Mpc^{-1})$, where $H_0$ is the Hubble parameter at the current
epoch.  It should be kept in mind that future calculations with more
sophisticated treatments of the singlet neutrino production physics
and the early universe environment may sharpen up, shift, or possibly
extend the mass/mixing parameter range which gives interesting relic
Dark Matter contributions.

For non-negligible lepton numbers, the singlet closure fraction
produced depends on the precise value of $L_{\nu_\alpha}$ as well as
the mixing angle and mass of the singlet neutrino.  In general, the
matter-enhanced ($L_{\nu_\alpha}\neq 0$) singlet neutrino production
modes can produce the same closure fraction as the
$L_{\nu_\alpha}\approx 0$ models but do so with orders of magnitude
smaller vacuum mixing angles (\citealt{Shi:1998km}; AFP). As a result,
it is in general harder to constrain the $L_{\nu_\alpha}\neq 0$
singlet neutrino production mode scenarios.

{
\refstepcounter{figure}
\centerline{\includegraphics[width=4.2cm]{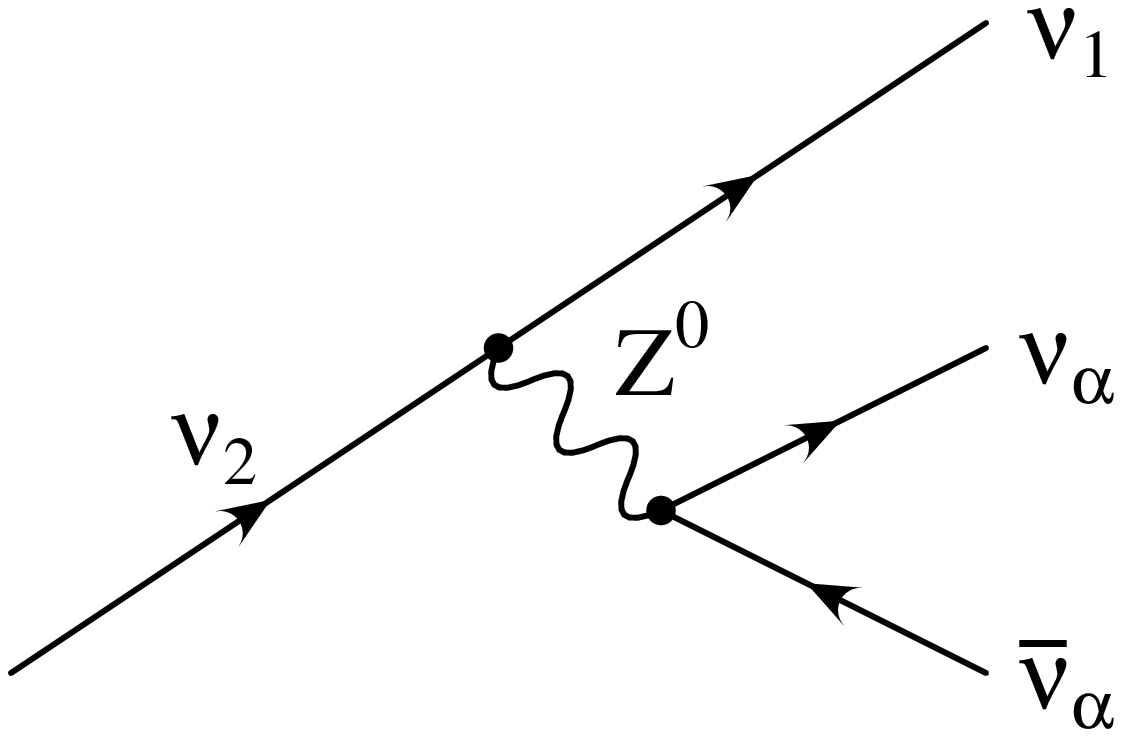}}
{\label{nunufeynman}
\footnotesize{FIG. 1 --- The principal decay mode for massive 
singlet neutrinos with mass less than twice the electron mass. There
are three light active neutrinos in the final state. 
(Here $\alpha = e,\mu,\tau$.)} }
\bigskip
}

Significant constraints can be made on massive neutrinos via the
effects of their decay \citep{Dicus:1977qy}. The primary decay channel
of massive singlet neutrinos is into three light active neutrinos and
is shown in Fig.\ \ref{nunufeynman}.  The decay rate corresponding to
this process is \citep{Barger:1995ty,BoehmVogel}
\begin{eqnarray}
\Gamma_{3\nu} &\approx& \sin^2 2\theta\ G_{\rm F}^2
\left(\frac{m_s^5}{768\pi^3}\right)
\label{decayrate}\\ &\approx& 8.7\times 10^{-31}
{\,\rm sec^{-1}} \left(\frac{\sin^2 2\theta}{10^{-10}}\right)
\left(\frac{m_s}{1{\rm\, keV}}\right)^5,\nonumber
\end{eqnarray}
where $G_F \approx 1.166\times 10^{-11}\rm\ MeV^{-2}$ is the Fermi
constant.  This process needs to be considered when comparing number
densities of singlets produced at a very early time with those today,
for example, in the calculation of the diffuse extragalactic
background radiation (see Section \ref{debrasect}).

{
\smallskip
\refstepcounter{figure}
\includegraphics[width=8.4cm]{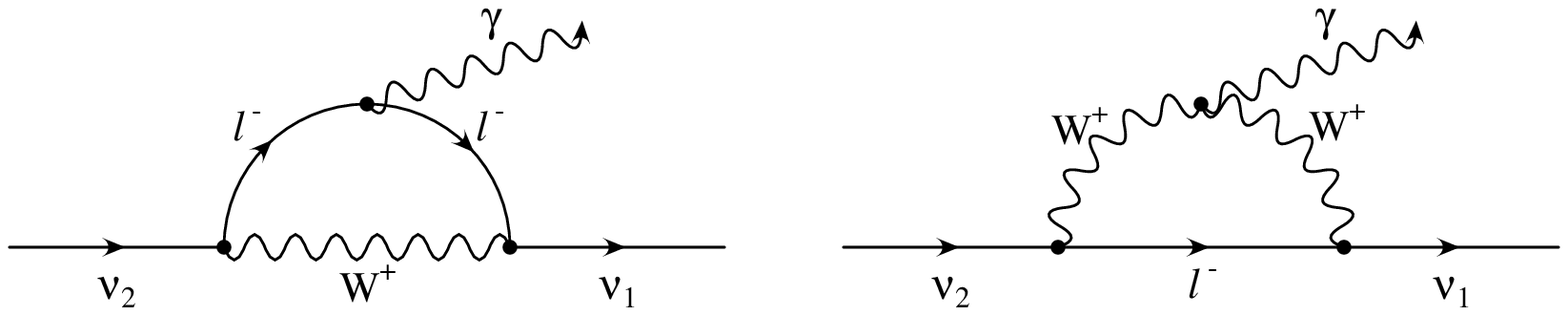}
{\label{nufeynman}
\footnotesize{FIG. 2 --- Principal radiative decay modes for massive 
singlet neutrinos.} }
\bigskip
}

The principal radiative decay modes of singlet neutrinos are shown in
Fig.\ \ref{nufeynman}. Majorana neutrinos have contributions from
conjugate processes.  For the Majorana neutrino case, the decay rate
for $m_2\gg m_1$, 
is \citep{Pal:1982rm}
\begin{equation}
\Gamma_\gamma = \frac{\alpha G_F^2}{64\pi^4} m_2^5 \left[\sum_\beta
U_{1\beta} U_{2\beta} F(r_\beta)\right]^2,
\label{sdecay}
\end{equation}
where $\alpha\approx 1/137$ is the fine structure constant.  Here,
$r_\beta = (m_{\beta}/M_W)^2$ is the square of the ratio of the
$\beta$ flavor charged lepton mass and the W$^\pm$ boson mass, and
\begin{equation}
F(r_\beta) \approx -\frac{3}{2}+\frac{3}{4} r_\beta.
\label{f}
\end{equation}
The sum in equation\ (\ref{sdecay}) is over the charged lepton flavors.
For decay of a doublet neutrino into another flavor doublet, the sum
in equation\ (\ref{sdecay}) vanishes for the first term in equation\ (\ref{f})
on account of the unitarity property associated with the
transformation matrix elements in equation\ (\ref{transformnu}).  The
second term in equation\ (\ref{f}) causes the sum not to vanish, but the
resulting term is obviously very small because it involves the fourth
power of the ratio of charged lepton to W$^\pm$ masses.  This is the
so-called Glashow-Iliopoulos-Maiani (GIM) suppression (or
cancellation).

For a singlet decay, the sum over the charged lepton flavors in equation\
(\ref{f}) does not cancel the leading contribution in equation\ (\ref{f})
because there is no charged lepton associated with the singlet
state. The decay rate is consequently greatly enhanced over the
GIM-suppressed doublet decay case.  The rate of singlet neutrino
radiative decay is
\begin{equation}
\Gamma_\gamma(m_s,\sin^2 2\theta) \approx 6.8 \times 10^{-33}\,{\rm
s^{-1}}\ \left(\frac{\sin^2 2\theta}{10^{-10}}\right)
\left(\frac{m_s}{1\,\rm keV}\right)^5,
\end{equation}
where we have identified $m_s \approx m_2$, since the mixing is
presumed to be small. 

The singlet neutrino can also decay via two-photon emission,
$\nu_2\rightarrow\nu_1+\gamma+\gamma$.  However, this decay has a
leading contribution scaling with the inverse square of the charged
lepton mass \citep{Nieves:1983bq}, and therefore is strongly
suppressed.  Since the two-photon decay rate scales as $m_s^9$, it
will dominate over the single photon mode for masses $m_s\gtrsim
10\rm\ MeV$.  However, singlet neutrino masses over 10 MeV are
excluded by other considerations (AFP).

In the case of the single photon channel, the decay of a
nonrelativistic singlet neutrino into two (nearly) massless particles
produces a line at energy $E_\gamma = m_s/2$ with a width given by the
velocity dispersion of the dark matter.  For example, clusters of
galaxies typically have a virial velocity dispersion of $\sim 300\rm\
km\ sec^{-1}$. Therefore, the emitted line is very narrow, $\Delta E \sim
10^{-3} E_\gamma$.  The observed width of the line will be given by
the energy resolution of the detector in this case.  For example, the
energy resolution of {\it Chandra}'s ACIS is $\Delta E\approx 200\rm\ eV$,
while the Constellation X project hopes to achieve a resolution of
$\Delta E\approx 2\rm\ eV$.

The luminosity from a general singlet neutrino dark matter halo is
[from equation\ (\ref{lumgen})]
\begin{equation}
{\cal L} \approx 6.1 \times 10^{32}{\rm \, erg\,sec^{-1}}\ 
\left(\frac{M_{\rm DM}}{10^{11}\,M_\odot}\right)
\left(\frac{\sin^2  2\theta}{10^{-10}}\right)
\left(\frac{m_s}{1\rm\, keV}\right)^5.
\end{equation}
This implies that the radiative decay flux from singlet neutrinos in
the halo is
\begin{eqnarray}
F \approx&& 
5.1\times 10^{-18} \flux
\left(\frac{D_L}{1\,\rm Mpc}\right)^{-2}
\left(\frac{M_{\rm DM}}{10^{11}\,M_\odot}\right)\cr
&&\times\left(\frac{\sin^2 2\theta}{10^{-10}}\right)
\left(\frac{m_s}{1\rm\, keV}\right)^5.
\label{fluxlimit2}
\end{eqnarray}
Therefore, for a general singlet neutrino candidate with rest mass $m_s$
and vacuum mixing angle $\sin^2 2\theta$, the mass limit---assuming no
detection of a line at a flux limit level of $F_{\rm det}$---is
\begin{eqnarray}
m_s \lesssim && 4.6\ {\rm keV} 
\left(\frac{D_L}{1\,\rm Mpc}\right)^{2/5}
\left(\frac{F_{\rm det}}{10^{-13}\flux}\right)^{1/5}\cr
&&\times\left(\frac{M_{\rm DM}}{10^{11}\,M_\odot}\right)^{-1/5}
\left(\frac{\sin^2 2\theta}{10^{-10}}\right)^{-1/5}.
\label{handy2}
\end{eqnarray}

Using equation\ (\ref{omega}), the dependence on mixing angle can be
eliminated, and with equation\ (\ref{lumgen}), we have for the
$L_{\nu_\alpha}\approx 0$ case that the flux due to singlet
neutrino decay is
\begin{eqnarray}
F \approx&& 
 5.1\times 10^{-14} \flux
\left(\frac{D_L}{1\,\rm Mpc}\right)^{-2}
\left(\frac{M_{\rm DM}}{10^{11}\,M_\odot}\right)\cr
&&\times\left(\frac{\Omega_{\nu_s} h^2}{0.3}\right)
\left(\frac{m_s}{1\rm\, keV}\right)^3.
\label{fluxlimit}
\end{eqnarray}
For the $L_{\nu_\alpha}\approx 0$ production case, the corresponding
singlet mass limit from a null detection of a line at $E_\gamma=m_s/2$
at flux limit $F_{\rm det}$ is
\begin{eqnarray}
m_s \lesssim && 1.25\ {\rm keV} 
\left(\frac{D_L}{1\,\rm Mpc}\right)^{2/3}
\left(\frac{F_{\rm det}}{10^{-13}\flux}\right)^{1/3}\cr
&&\times\left(\frac{M_{\rm DM}}{10^{11}\,M_\odot}\right)^{-1/3}
\left(\frac{\Omega_{\nu_s} h^2}{0.3}\right)^{-1/3}.
\label{handy}
\end{eqnarray}

It should be noted that the decay limits presented here derive from a
specific type of mass-generation mechanism for the singlet neutrino:
those arising from the simplest case of Majorana or Dirac type mass
terms.  More complicated neutrino mass models would have different
mass-terms, radiative decay widths and possibly other couplings, and
bounds on these models would require individual analysis.

\subsection{Active Neutrinos}

The direct experimental upper limits on the $\nu_\mu$ and $\nu_\tau$
masses are only $190\rm\ keV$ and $18.2\rm\ MeV$, respectively
\citep{Groom:2000in}.  Although the observationally-inferred age of
the universe precludes the possibility of fully thermalized active
neutrinos being the WDM or CDM \citep{Gerstein1966,Cowsik:1972gh}, the
active neutrinos may not be fully thermalized in the early universe if
the post-inflation reheating temperature is low
\citep*{Kawasaki:2000en,Giudice:2000ex}.  In this case, the $\nu_\mu$
and/or $\nu_\tau$ can be legitimate WDM candidates
\citep{Giudice:2000dp}.

The radiative decay rate of massive active neutrinos into a single
photon in a three-generation neutrino model is suppressed by the GIM
mechanism, and is slower than the singlet neutrino decay rate by a
factor $(m_{\alpha}/M_W)^4\sim 10^{-21}$ and is therefore negligible.
The two photon decay mode $\nu_2\rightarrow\nu_1+\gamma+\gamma$ for
massive active neutrino decay has a GIM suppression factor of $\sim
(m_{a}/m_{\alpha})^4$ where $m_a$ is the relevant vacuum neutrino mass
eigenvalue, and $m_\alpha$ is the mass of the charged lepton of flavor
$\alpha$ \citep{Nieves:1983bq}. This rate can therefore dominate the
single photon mode only for $m_a \gtrsim 200\rm\ keV$, which is
outside of the range of interest for active neutrino WDM.  Therefore,
active neutrino WDM in a three-generation neutrino model is relatively
stable and robust against decay constraints.

In order to accommodate solar and atmospheric neutrino oscillation
solutions to experimental data, a light singlet neutrino with mass
$m_s\ll m_{\nu_{\mu,\tau}}$ must be a feature of any neutrino
mass/mixing scheme which can provide for an active $\nu_\mu/\nu_\tau$
WDM candidate.  If there is a light singlet, then the decay from the
mass eigenstates most closely associated with the $\nu_\mu/\nu_\tau$
doublet into a mass state more closely associated with the lighter
singlet is not GIM suppressed. This process can lead to a rapid
radiative decay of the active WDM candidate.  However, the mixing
between the two mass eigenstates can be tuned to be arbitrarily small,
since it is not phenomenologically required to be non-zero.  If a
spectral feature in the x-ray is observed, however, it may indicate a
non-negligible mixing between a massive active neutrino and a lighter
singlet neutrino.

\subsection{Gravitinos}

Another WDM candidate is the gravitino, $\tilde{G}$, the spin-1/2
supersymmetric partner to the graviton \citep*{Kawasaki:1997wc}.  In
the minimal supersymmetric standard model the assumption of R-parity
conservation is made, originally motivated by the need to explain the
slow proton decay rate.  The gravitino can be the lightest
supersymmetric particle (LSP) and, with R-parity conservation, it is
stable.  However, there has been considerable interest in R-parity
violation in supersymmetric models as a mechanism for neutrino mass
generation \citep{Hall:1984id}, motivated by the considerable evidence
for neutrino mass [for a review, see \citet{Caldwell:1998fq}].  The
gravitino may also be the LSP and the dark matter
\citep{Takayama:2000uz} in R-parity violating models with gauge
mediation \citep{Dine:1996ag} or with a low gravitational scale (or
effective Planck scale $M_{\rm Pl}$) which comes about through large
extra dimensions \citep*{Arkani-Hamed:1998rs}.

The gravitino LSP decays with a long lifetime in these models since
its decay is suppressed by $M_{\rm Pl}^2$.  In one specific example
considered by \citet{Takayama:2000uz}, the R-parity violation is
bilinear with the lightest neutralino being bino-dominant.  In this
case, the dominant decay mode of the gravitino is
$\tilde{G}\rightarrow\gamma\nu$ through a coupling of the gravitino
with the photon and its superpartner, the photino, which has a
neutrino component.  The lifetime of the gravitino is approximately
\begin{equation}
\Gamma_\gamma(\tilde{G}\rightarrow\gamma\nu) \approx \frac{1}{4}
|U_{\gamma\nu}|^2 \frac{m_{3/2}^3}{M_{\rm Pl}^2},
\end{equation}
where $U_{\gamma\nu}$ represents the neutrino component of the
photino, and $m_{3/2}$ is the gravitino mass. 

In the models accomodating neutrino masses associated with the
atmospheric neutrino problem, $U_{\gamma\nu}$ has a characteristic
value $|U_{\gamma\nu}|^2 \approx 7\times 10^{-13}$.  The decay rate of
the gravitino into photons is then
\begin{eqnarray}
\Gamma_\gamma(\tilde{G}\rightarrow\gamma\nu) &\approx& (2.6 \times
10^{19}{\rm\ yr})^{-1}\ \left(\frac{m_{3/2}}{1\rm\ GeV}\right)^3
\left(\frac{|U_{\gamma\nu}|^2}{7\times 10^{-13}}\right)\nonumber\\
&&\times \left(\frac{M_{\rm Pl}}{M^0_{\rm Pl}}\right)^{-2},
\label{gravitinorate}
\end{eqnarray}
where $M^0_{\rm Pl}\approx 1.22\times 10^{19}\rm\ GeV$ is the
conventional Planck scale.  The decay would produce a photon line at
an energy of $E_\gamma\approx m_{3/2}/2$. For a $\sim$1 keV mass
gravitino, the decay rate is far below the detectable limit given by
equation (\ref{decaylimit}).  However, as mentioned previously, the
gravitino is the LSP in some supersymmetric models with large extra
dimensions.  These scenarios generically reduce the effective Planck
scale by up to 14 orders of magnitude.  From
equation (\ref{gravitinorate}), if the effective Planck scale is reduced by
``only'' 7 orders of magnitude, the rate becomes detectable.
Ultimately, the lack or presence of a photon emission line may
constrain supersymmetric dark matter models with large extra
dimensions.

\section{Observing Dark Matter Halos}

Astronomical objects with strong evidence for dark matter
concentrations can serve as source reservoirs for WDM particles. The
specific amount of dark matter in the observed region can be inferred
from models of the spatial distribution of the dark matter.  The
theoretical basis for the dark matter distribution is either based on
local or cosmological physics.

Local models include truncated isothermal sphere configurations, which
describe galactic dark matter halos as perfect gasses in equilibrium.
The isothermal sphere model reproduces well the observed flat rotation
curves of spiral galaxies. In another local-type model, the dark
matter in x-ray clusters of galaxies is assumed to be the dominant
source of the gravitational potential binding the hot x-ray emitting
intracluster medium in hydrostatic equilibrium in the $\beta$-model.

In CDM models, both galaxy and galaxy cluster halos can be the
result of hierarchical clustering and so can possess a unified
profile.  However, CDM models and their corresponding unified profiles
may be too centrally concentrated to describe dwarf galaxy rotation
curves \citep*{Navarro:1996iw,Navarro1995}.  The general dark matter
density profile with radius $\rho(r)$ for hierarchical clustering is
dubbed the Navarro-Frenk-White (NFW) profile:
\begin{equation}
\rho(r) \propto \left(\frac{r}{r_s}\right)^{-1}
\left(1+\frac{r}{r_s}\right)^{-2},
\end{equation}
where $r_s$ is the scale radius. 

Given any dark matter distribution for an object, the photon flux
resulting from particle decay or interaction is simply proportional to
the mass within the field of view.  This can be approximated as the
mass within the projected radius at the distance of the object.
However, the dark matter in the field of view contains not only the
extended halo out to the projected radius but also the material in
front of and behind this sphere.  To determine the mass within the
rectangular prism cut out of the extended halo of the object, we have
performed a Monte Carlo integration of the region's mass density,
sampling according to an assumed density profile, e.g., NFW.

\subsection{Instrumental Background}

For low surface brightness dark matter systems, the primary limit to
dark matter flux detection is instrumental background.  In particular,
ACIS aboard {\it Chandra} has a background of $2\times 10^{-2}\rm cts\
sec^{-1}$ in a $\sim$200 eV energy bin of the imaging array.

As a rough approximation, we can estimate the flux onto ACIS from dark
matter decay requisite to produce a 4-$\sigma$ detection. For a line
with flux $F_{-14}\equiv F/(10^{-14} \flux)$, the count rate is $C_L
\approx 3\times 10^{-4} F_{-14}\rm\ cts\ sec^{-1}$.  For an
observation of integration time $t_5\equiv t/(10^{5}\rm\ sec)$, the
background is $B\approx 2\times 10^3 t_5\rm\ cts$. The count level
required to overcome the background at 4-$\sigma$ is
\begin{equation}
C_L \approx \frac{4\sqrt{B}}{t}.
\end{equation}
Therfore, the detectable flux is 
\begin{equation}
F^{\rm det}_{-14} \approx 6\ t_5^{-1/2}.
\end{equation}
For low surface brightness sources, an observation with an integration
time of 36,000 sec ($t_5=0.36$) can detect a flux $\approx
10^{-13}\flux$ ($F^{\rm det}_{-14} =10$). One must modify the
estimated detectable WDM flux whenever the baryon-associated material
(e.g., electrons, protons, iron, etc.) in these halos provides
an ambient x-ray flux in excess of $3\times 10^{-11} \flux$.  In such
circumstances the minimum detectable WDM flux in a $\sim$200 eV energy
bin, corresponding to the ACIS energy resolution, is
\begin{equation}
F^{\rm det}_{-14} \approx 6\ \left(\frac{F_{\rm source}}{3\times
10^{-11}}\right)^{1/2} t_5^{-1/2}\flux .
\end{equation}

{
\bigskip
\refstepcounter{figure}
\centerline{\includegraphics[width=9cm]{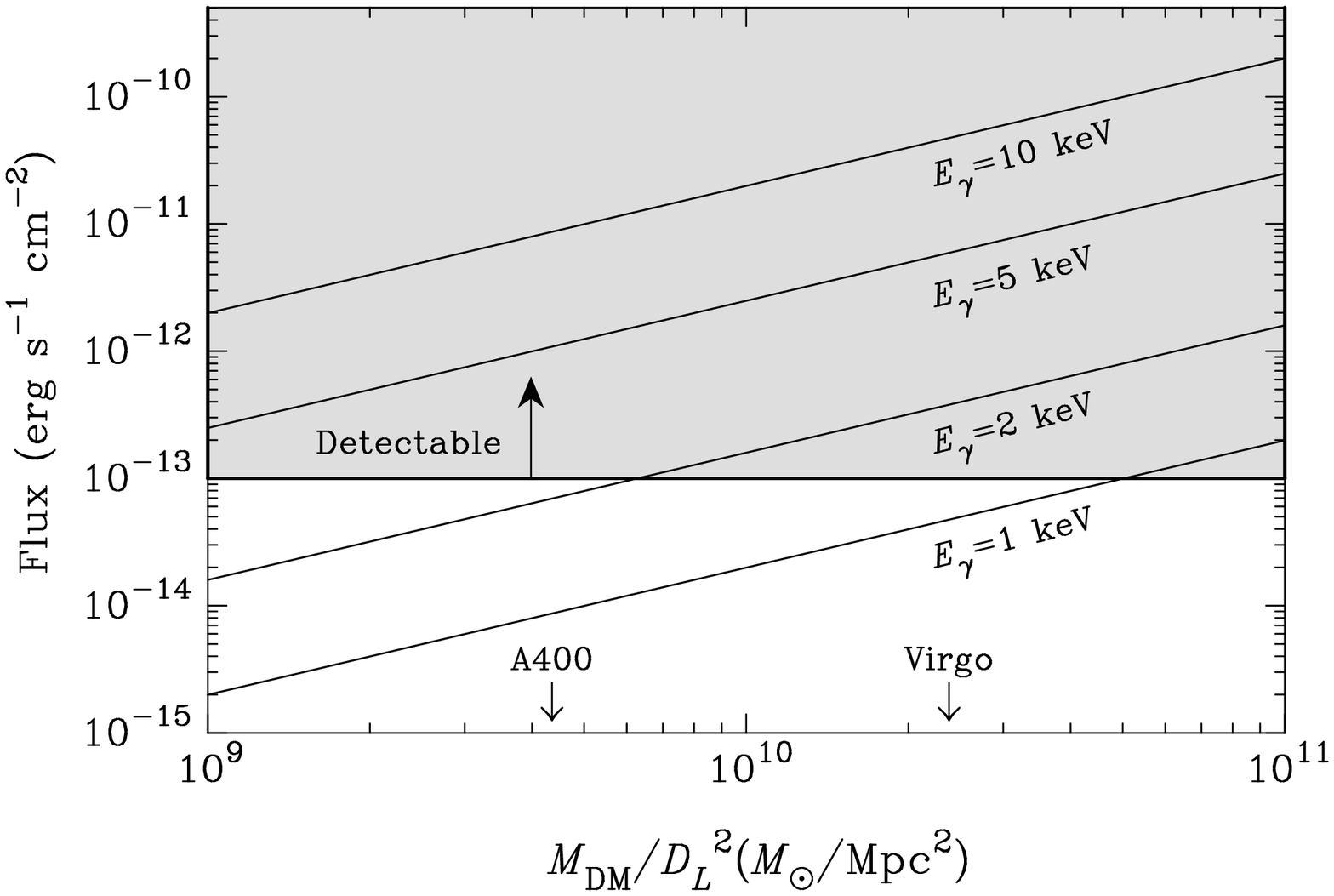}}
{\label{massvsrc}
\footnotesize{FIG. 3 --- Singlet neutrino decay x-ray flux produced by an
object with a mass $M_{\rm DM}$ at distance $D_L$.  Shown are contours
that correspond to spectral lines with energy $E_\gamma$ produced by singlet
neutrinos of mass $m_s=2 E_\gamma$.  The shaded region indicates
detectability with ACIS on {\it Chandra} for a 36 ksec observation.}}
\bigskip
}

In Fig. \ref{massvsrc} we show contours of WDM x-ray photon decay flux
in lines at photon energies $E_\gamma=m_s/2$ as a function of the dark
matter mass in the field of view modulo its luminosity distance [see
equation\ (\ref{fluxlimit})], for a WDM singlet neutrino model with zero
initial lepton number, $L_{\nu_\alpha}\approx 0$.  This figure
provides a shaded region indicating detectability with ACIS on
{\it Chandra} for a 36 ksec observation.

\subsection{Clusters of Galaxies}

Clusters of galaxies are massive objects ($\sim\! 10^{15}M_\odot$) and
there is strong evidence that most of this mass is carried by dark
matter.  Mass estimates are derived from a spherical hydrostatic
equilibrium model that assumes the gas in the intracluster medium is
solely supported by thermal pressure.  The enclosed mass at a radius
$r$ is given by
\begin{equation}
M(<r) = - \frac{k T_{\rm X}(r)}{G \mu m_p} r \left[\frac{d\, \log
 \rho_g(r)}{d\, \log r} + \frac{d\, \log T_{\rm X}(r)}{d\, \log r}\right],
\end{equation}
where $G=(M^0_{\rm Pl})^{-2}$, $\mu m_p$ is the average molecular
weight of the gas ($m_p$ is the proton mass), $k$ is Boltzmann's
constant, $\rho_g(r)$ is the gas density profile, and $T_{\rm X}(r)$ is the
temperature profile. Using this equilibrium model, an approximate
isothermal fit to the mass profile can be made.  This is the so-called
isothermal $\beta$-model.  The enclosed mass for the
isothermal $\beta$-model is
\begin{equation}
M(<r) \approx 1.13\times 10^{14}\,M_\odot\left(\frac{T_{\rm X}}{\rm keV}\right)
\left(\frac{r}{\rm
Mpc}\right)\frac{(r/r_c)^2}{1+(r/r_c)^2},
\end{equation}
where $r_c$ is the core radius \citep{Cavaliere1978}.  (Note that
$r_c$ and $r_s$ are related but not identical.)

With contemporary x-ray telescopes (ASCA, {\it XMM-Newton} and {\it Chandra}) able
to provide spatially resolved temperature profiles, the hydrostatic
equilibrium $\beta$-model in principle can yield accurate mass
estimates.  However, as yet, detailed analyses along these lines to
provide $T_{\rm X}(r)$ are still being done.  Therefore, we employ the
isothermal $\beta$-model with average temperature values.  These are
available for the 24 clusters we consider.

In this work we use the average temperatures of rich clusters from the
\citet*{Horner1999} database. Redshifts or distances are taken from
the SIMBAD database.\footnote{Available at
http://simbad.u-strasbg.fr.}  The physical properties of the clusters
we consider are given in Table 1.

We can approximate the reservoir of dark matter mass and observable
decay flux for singlet neutrino dark matter seen by ACIS aboard the
{\it Chandra} observatory using the relation between the projected radius of
the field of view, $R_{\rm fov}$, and the angular field of view of
ACIS, $\theta_{\rm fov} \approx 5\times 10^{-3}\rm\ rad$: $D_L = 2
R_{\rm fov}/\theta_{\rm fov} \approx 4\times 10^2 R_{\rm fov}$.  With
equation\ (\ref{fluxlimit}), for the $L_{\nu_\alpha}\approx 0$ case, the
expected x-ray flux is then
\begin{eqnarray}
F &\approx& 3\times 10^{-19}\flux \left(\frac{m_s}{\rm
keV}\right)^3 \left(\frac{\Omega_{\nu_s} h^2}{0.3}\right)\cr
&&\times\left(\frac{M_{\rm DM}}{10^{11}\,M_\odot}\right)
\left(\frac{R_{\rm fov}}{1\,\rm Mpc}\right)^{-2}.
\end{eqnarray}

The core radius falls within the field of view of {\it Chandra} for all
clusters we consider except for the Coma Cluster.  Since the mass
increases approximately with the radius outside of the core, we can
approximate $M_{\rm fov} \sim M_{\rm core} R_{\rm fov}/r_c$.
Therefore, the expected x-ray flux from the cluster in the
$L_{\nu_\alpha}\approx 0$ model is
\begin{eqnarray}
F &\approx& 3\times 10^{-19} \flux \left(\frac{m_s}{\rm keV}\right)^3
\left(\frac{M_{\rm core}}{10^{11}\,M_\odot}\right)\cr
&&\times\left(\frac{r_c}{1\,\rm Mpc}\right)^{-1}
\left(\frac{R_{\rm fov}}{1\,\rm Mpc}\right)^{-1}\cr
&\approx&1.2\times 10^{-16} \flux \left(\frac{m_s}{\rm keV}\right)^3
\left(\frac{M_{\rm core}}{10^{11}\,M_\odot}\right)\cr
&&\times\left(\frac{D_L}{1\,\rm Mpc}\right)^{-1}
\left(\frac{R_{\rm fov}}{1\,\rm Mpc}\right)^{-1}.
\end{eqnarray}

In order to more accurately determine the dark matter mass and
luminosity in decay photons, we estimate the dark matter mass within
the rectangular prism cut out of the spherical isothermal
$\beta$-model density distribution in the dark matter halo.  The prism
has the dimensions of $2R_{\rm fov}\times 2R_{\rm fov}\times 2 R_{\rm
vir}$, where $R_{\rm fov}$ is half the projected size of the
aperture, $R_{\rm fov}\approx \theta_{\rm fov} D_L/2$, and where
$R_{\rm vir}$ is the cluster's virial radius.  We use a Monte Carlo
integration to find the dark matter mass within this field of view,
which can be significantly larger than the mass strictly within the
spherical volume defined by the core radius.

We can use equation\ (\ref{handy}) to provide a rough limit on the mass
$m_s$ of a singlet neutrino candidate assumed to comprise the dark
matter halo.  The limits on $m_s$ for various clusters are given in
Table 1.  The best limit, $m_s \lesssim 2.6\rm\ keV$ (for the
$L_{\nu_\alpha}\approx 0$ case), is provided by the Virgo Cluster.

{
\bigskip
\refstepcounter{figure}
\centerline{\includegraphics[width=9cm]{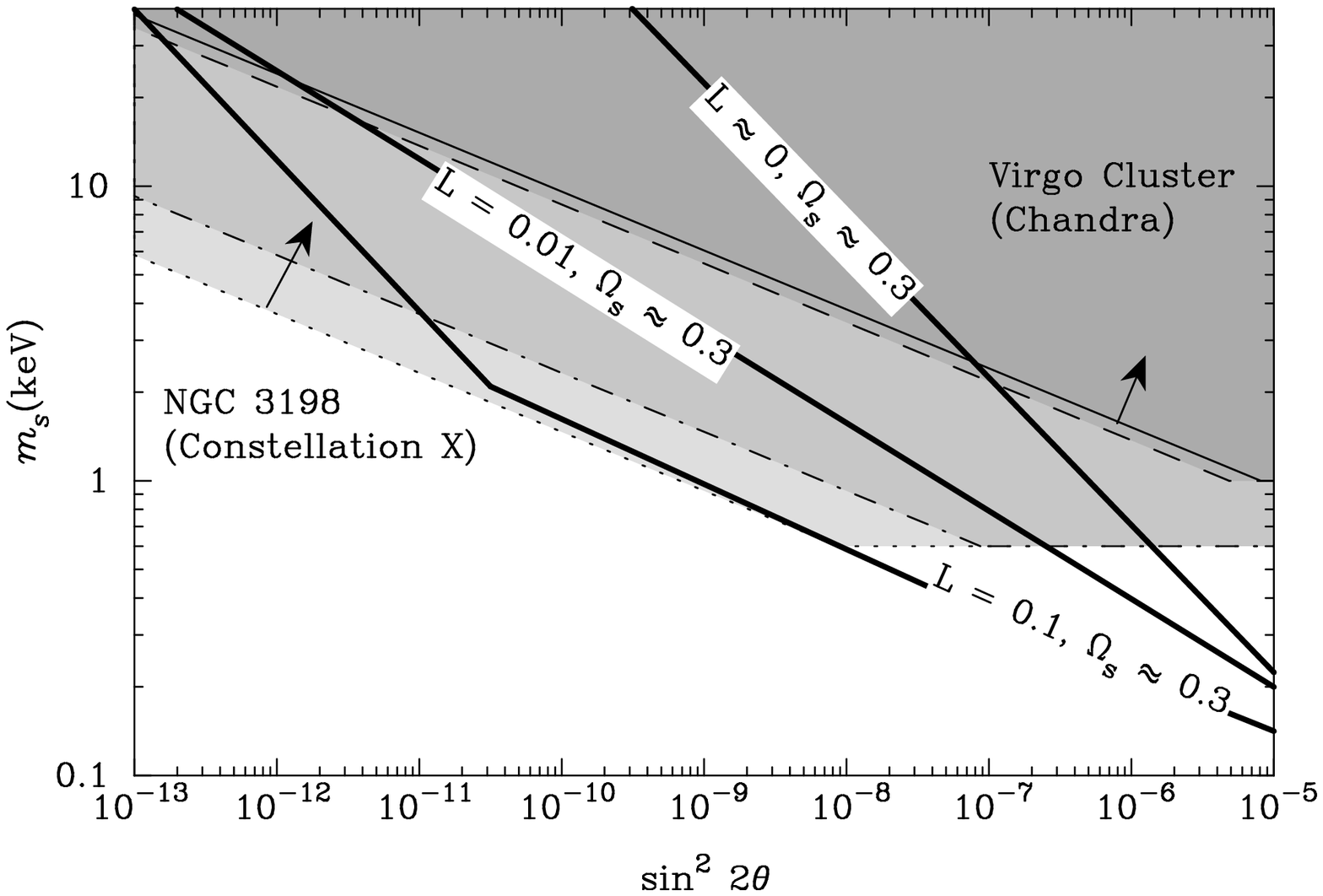}} {\label{virgo}
\footnotesize{FIG. 4 --- Shown are approximate contours for $\Omega_s
\approx 0.3\ (h=0.7)$ for singlet neutrino WDM models with different
initial lepton number, $L$.  Potential detection/exclusion of the
models can be made in the darkest gray region with a limiting line
flux of $10^{-13}\flux$ from the Virgo Cluster observed with {\it Chandra}'s
ACIS (from a 36 ksec observation).  The dashed line is the limit for a
100 ksec observation with {\it Chandra} also for Virgo. The medium gray
region bounded by the dashed-dotted line indicates the
detection/exclusion range for Constellation X observation of field
spiral galaxy NGC 3198 for a 1000 ksec observation, and the light gray
region indicates possibilty of detection/exlusion for an ambitious 10
Msec observation (presumably obtained in several observations over a
few years).}}
\bigskip
}

In Fig.\ \ref{virgo} we give contours in singlet mass $m_s$ and vacuum
mixing angle space of singlet neutrino closure fraction
$\Omega_s\approx 0.3$ (for $h=0.7$) for three values of primordial
lepton number ($L\equiv L_{\nu_\alpha}$): $L\approx 0, L\approx 0.01,
L\approx 0.1$.  The non-zero lepton number contours are very rough
fits to the results of AFP. (Note that lepton number $L$ here is denoted
as $\cal L$ in AFP.) Superimposed on this figure are shaded
regions indicating potential detectability or exclusion of singlet
neutrino WDM from {\it Chandra} observations of the Virgo Cluster for a 36
ksec observation (dark shade bounded by the solid line) and a 100 ksec
observation (bounded by the dashed line).  The {\it Chandra} observations of
the Virgo Cluster can come close to eliminating all of the $L\approx
0$, non-resonant singlet production mode case.  In the next section,
we will argue that the Constellation X project can give even broader
constraints in this figure.

To study the possible observational signature of a singlet neutrino
halo in the Virgo Cluster, we generate spectra of the gas in Virgo as
seen with ACIS with WEBSPEC (XSPEC).  We use the MEKAL model for the
x-ray flux for the gas at a temperature of 2.54 keV \citep{Horner1999}
with a thermal x-ray flux from the central region of $1.5\times
10^{-12}\flux$ in the 2-10 keV band \citep{Bohringer2001}.  We adopt a
distance to the Virgo cluster of 20.7 Mpc measured from 21 cm line
widths by \citet*{Federspiel1998}.   

\begin{figure*}
\centerline{\includegraphics[width=14cm]{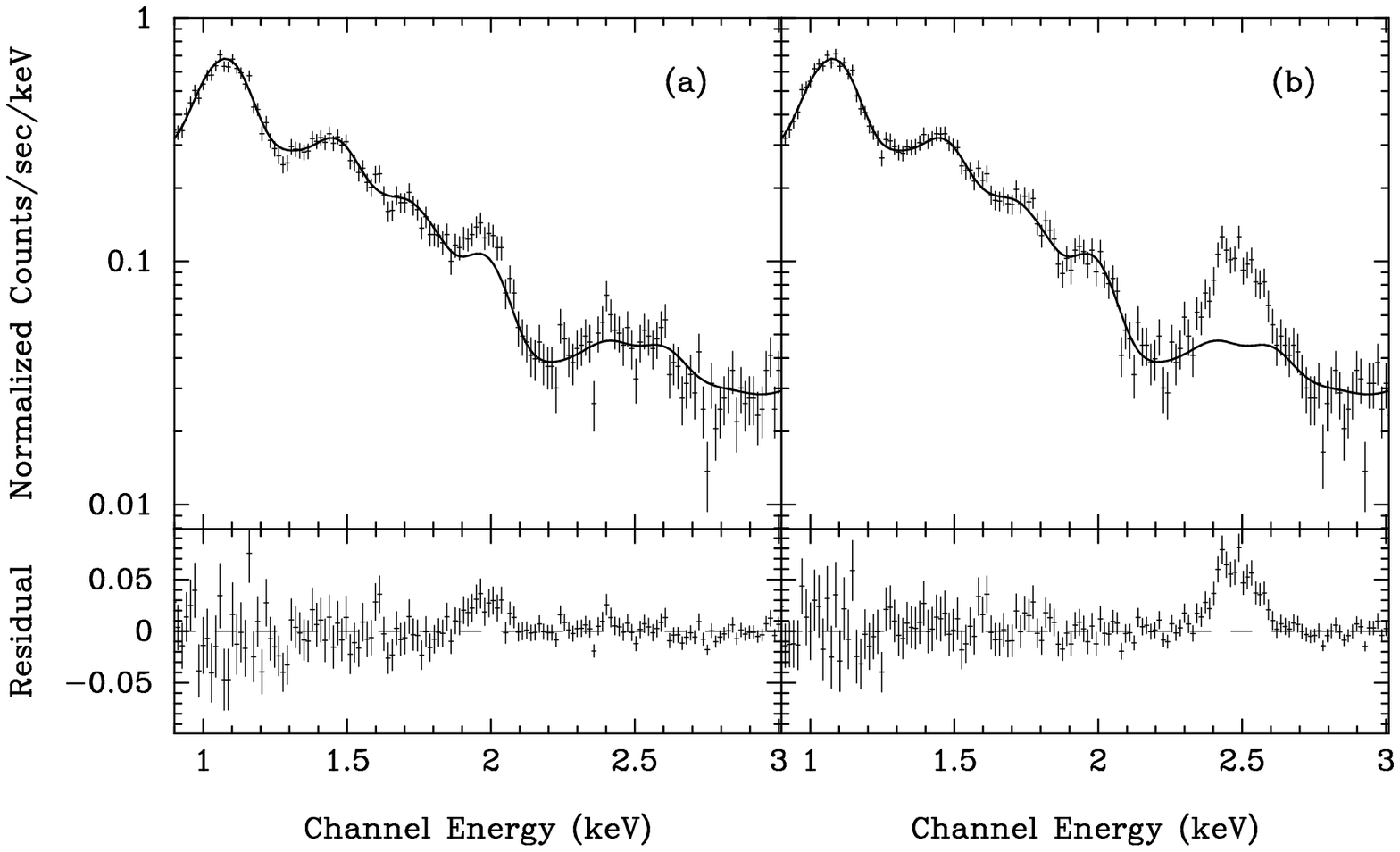}}
\caption{
\label{spectrum}
\footnotesize{A synthesized spectrum viewed by {\it Chandra}'s ACIS modeling
the central region of the Virgo cluster, for two different cases of
singlet neutrino mass, showing a strong mass dependence of the flux in
a line at (a) a 4 keV singlet neutrino halo (producing a 2 keV line),
and (b) a 5 keV singlet neutrino halo (producing a 2.5 keV line).
Integration time is assummed to be 50 ksec, with a $T_{\rm X} = 2.54\rm\
keV$ thermal flux ($1.5 \times 10^{-12}\flux$) from the gas in Virgo
generated with the MEKAL model, which is shown as the solid line.
Residuals from the gas emission model alone are shown at the bottom.
The width of the line is nearly entirely due to instrumental
energy resolution.}}
\end{figure*}

In Fig.\ \ref{spectrum}, we show the binned spectrum for two cases. We
include the decay fluxes [see equation\ (\ref{fluxlimit})] from a singlet
neutrino halo composed of 4 keV and 5 keV singlet neutrinos producing
decay photons of energy 2 and 2.5 keV, respectively, for a 50 ksec
observation on {\it Chandra}'s ACIS.  In addition, we include the
theoretical MEKAL model gas emission spectrum. The residuals from this
standard gas emission prediction are given at the bottom of the
figure.  The width of the lines are nearly entirely due to
instrumental broadenning.

From this unsophisticated example, it is obvious that a line feature
produced by a 5 keV singlet neutrino would be readily detectable,
while the detection/elimination of the decay line for a 4 keV singlet
would require a statistical analysis.  With the lack of such a strong
line anomaly in the observation of M87 in the core of the Virgo
Cluster with {\it XMM-Newton} by \citet{Bohringer2001}, we can conclude that
the dark matter in the Virgo Cluster is not composed of singlet
neutrinos with masses $m_s \gtrsim 5\rm\ keV$ created in the early
universe with $L_{\nu_\alpha} \approx 0$.  The potential constraints
on general $L_{\nu_\alpha}$ scenarios from statistical spectral
analyses are shown in Fig.\ \ref{virgo}.

\subsection{Field Spiral Galaxies}
\label{fieldgal}

Field galaxies can provide low gas x-ray emission sources for dark
matter, but these objects do not represent as concentrated a dark
matter source as do clusters of galaxies.  A detailed observation of
the structure and rotation of NGC 4123 has allowed a fit to disk and
halo models that places a lower limit on the profile of the dark
matter halo \citep{Wiener2001a}.  Using a maximal disk model to fit
the observed rotation curve, they find that a dark halo remains
required.  We use an NFW-type profile halo fit by \citet{Wiener2001a},
and find the resulting lower bound on the dark matter mass in the
field of view.  The distance to NGC 4123 is approximately 22.4
Mpc. The form of the NFW profile used is the two-parameter spherical
density distribution
\begin{equation}
\rho_{\rm NFW} = \rho_s \frac{4 r_s^3}{r(r+r_s)^2}.
\end{equation}
The halo parameters are from \citet{Wiener2001a} and are given in
Table 2.  The possible limit from a 36 ksec exposure [see equation\
(\ref{handy})] is $m_s \lesssim 10.3\rm\ keV$ in the
$L_{\nu_\alpha}\approx 0$ production mode case. 

We also use the rotation curve of NGC 3198 \citep{vanAlbada1985} to
fit an NFW profile for the dark matter in this object.  We use a
mass-to-light ratio of unity, which provides a good fit for the inner
part of the rotation curve.  The halo parameters are given in Table 2.
We adopt a distance of 18.34 Mpc to NGC 3198, from
\citet{Willick2001}. The potential mass limit [equation\ (\ref{handy})] for
a 36 ksec observation in this case is $m_s \lesssim 6.0\rm\ keV$ for
the $L_{\nu_\alpha}\approx 0$ models.  

In Fig.\ \ref{virgo} we show the detectability region for observations
of NGC 3198 with Constellation X---a proposed fleet of observatories
that will have an effective area $\sim$10 times greater than {\it Chandra},
and no instrumental background \citep{Valinia1999}---for two
integration times, 1 Msec and 10 Msec, which conceivably could be
achieved through several long observations over a few years.  An
exposure equivalent to this could be obtained by a stacking analysis
of the spectra of a number of similar clusters (see,
e.g. \citealt{Brandt2001} and \citealt{Tozzi2001}).  Constellation X,
with very long integration times, holds out the prospect of covering
nearly the entire WDM parameter space of interest for some of the
resonant production mode scenarios up to lepton number $L\lesssim
0.1$.

We should note, however, that it is still uncertain how low in $m_s$
(i.e., x-ray photon line energy) Constellation X can be
sensitive to at the limiting x-ray flux given in Fig.\ \ref{virgo}
($10^{-19}\flux$).  In fact, it could be that the lowest photon energy
detectable on Constellation X could be between 0.3 and 0.5 keV. We
show the possible constraint with detectability down to 0.3 keV
photons, or $m_s\approx 0.6\rm\ keV$.  As described in the
introduction, Lyman-$\alpha$ forest considerations disfavor
$m_s\lesssim 2.0\rm\ keV$, and therefore Constellation X in principle
can detect/exclude essentially all of the remaining
singlet neutrino parameter space for $L_{\nu_\alpha}\lesssim 0.1$.

\section{Diffuse Photon Limits}
\label{debrasect}

The flux per unit energy per unit solid angle from a homogenously
distributed decaying background dark matter particle is
\citep{Masso:1999wj}
\begin{equation}
\frac{d^2F}{dE_\gamma\ d\Omega} = \frac{\Gamma_\gamma}{4\pi}\
\frac{\tilde{n}_{\nu_s}(t_0)}{H(z_0)} e^{-\Gamma_{\rm tot} t(z_0)},
\label{difflux}
\end{equation}
where $\tilde{n}_{\nu_s}(t_0)$ is the present number density of dark
matter if it did not decay, $\Gamma_{\rm tot}$ is the total decay rate
of the particle, $z_0$ is the redshift at which the photon was
produced, and $t(z_0)$ is the age of the universe at $z_0$.  A photon
that has present energy $E_\gamma$ was produced at redshift $z_0$
given by
\begin{equation}
1+z_0 = \frac{m_s/2}{E_\gamma}.
\end{equation}

\begin{figure*}
\centerline{\includegraphics[width=16cm]{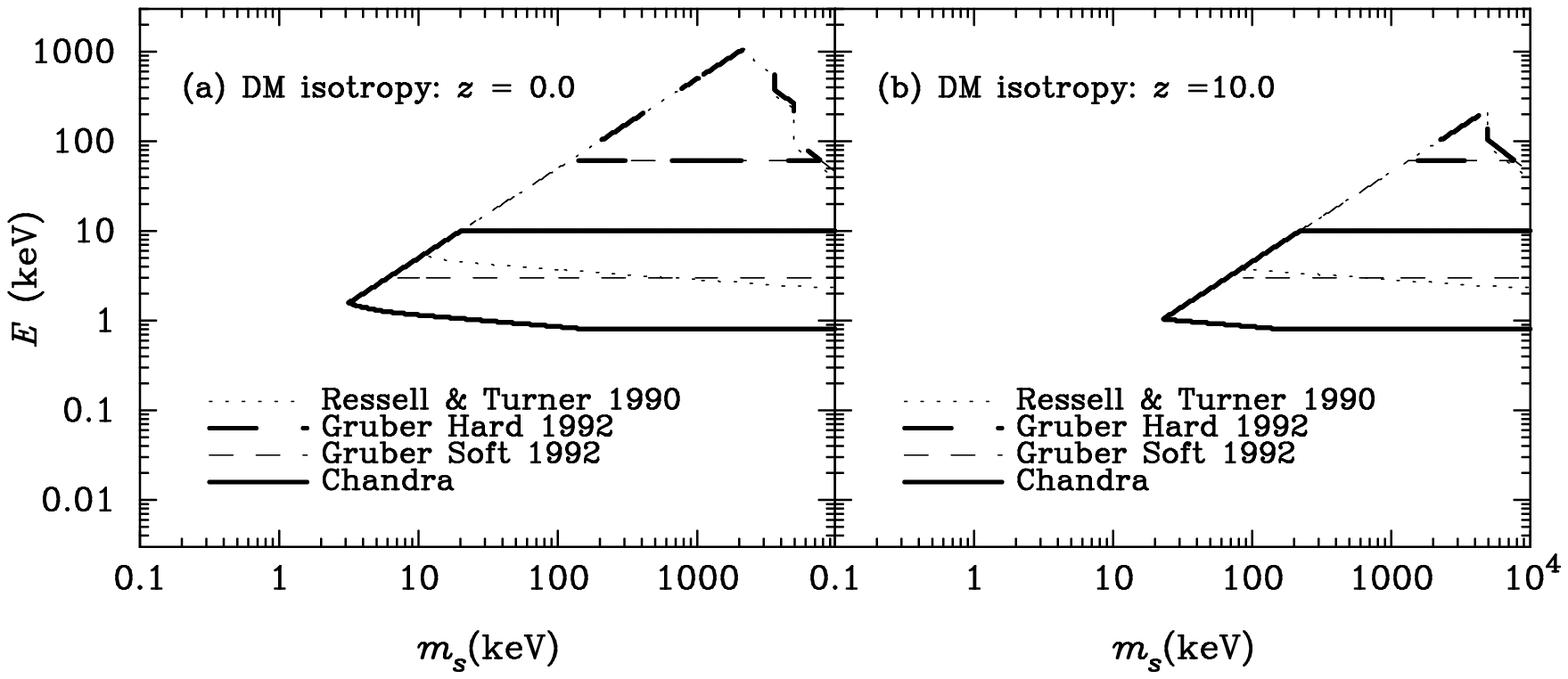}}

\caption{\label{diffuse} \footnotesize{ Limits on the diffuse flux
from singlet neutrino decay in the $L_{\nu_\alpha}\approx 0$ case from
photons produced at (a) $z\geq 0$ and (b) $z\geq 10$, assuming
isotropy of the dark matter at that redshift.}}
\end{figure*}

Limits on the diffuse extragalactic background radiation (DEBRA)
[sometimes referred to as extragalactic background light (EBL)] can be
used to constrain dark matter particles that produce a flux given by
equation (\ref{difflux}).  Diffuse radiative decay constraints also were
emphasized by \citet{Drees:2000qr} and \citet{Drees:2000wi} in papers
which incorrectly estimate the relic density of singlet neutrinos (see
\citealt{Dodelson:1994je,Dolgov:2000ew} and AFP). A broad-band limit
was placed by \citet{Ressell1990} on DEBRA. They found that the flux
per unit solid angle must satisfy
\begin{equation}
d{\cal F}/d\Omega \lesssim
(1{\rm\,MeV}/E_\gamma)\rm\ cm^{-2}sr^{-1}s^{-1}.\label{debralimit}
\end{equation}
Shown in Fig. \ref{diffuse} are these less constraining, yet
broad-band, bounds from \citet{Ressell1990} for the
$L_{\nu_\alpha}\approx 0$ case for singlet neutrino WDM.

\citet{Gruber1992} found the form of the x-ray background 
to be
\begin{equation}
d{\cal F}/d\Omega \lesssim 7.9 {\rm\ cm^{-2}sr^{-1}sec^{-1}}
\left(\frac{E_\gamma}{\rm keV}\right)^{-0.29} e^{-(E_\gamma/41\rm\ keV)}
\end{equation}
for energies of $\sim\! 3-60 \,\rm keV$, and 
\begin{eqnarray}
d{\cal F}/d\Omega &\lesssim& 1650 {\rm\ cm^{-2}sr^{-1}sec^{-1}}\ 
\left(\frac{E_\gamma}{\rm\ keV}\right)^{-2.0}\cr &&+1750 {\rm\
cm^{-2}sr^{-1}sec^{-1}}\  \left(\frac{E_\gamma}{\rm\
keV}\right)^{-0.7}
\end{eqnarray}
from $60\rm\, keV$ to $\sim\! 6 \rm\, MeV$.  These constraints are
also shown in Fig.\ \ref{diffuse} for the $L_{\nu_\alpha}\approx 0$
singlet WDM case.

Recently, the {\it Chandra} Observatory \citep{Hornschmeier2001,Tozzi2001}
has resolved several structures that contributed to the unresolved
x-ray background.  They find that the resolved sources contributed
from 60\% to 90\% of the previously unresolved backround.  We make a
rough approximation from these considerations that at most $\sim$20\%
of the x-ray background described by \citet{Gruber1992} may be due to
a diffuse particle decay.  The limits from this consideration are
depicted in Fig.\ \ref{diffuse} for the $L_{\nu_\alpha}\approx 0$
case. 

However, these diffuse limits entail the implicit assumption that dark
matter particles are not concentrated in objects, but rather are {\it
uniformly} distributed through space.  This is certainly not the case,
since structure is at least somewhat clumped and non-isotropic in the
sky since an epoch corresponding to a redshift of $z\gtrsim 10$.  The
limits on the singlet neutrino mass in the $L_{\nu_\alpha}\approx 0$
case obtained by an assumption of no clumping right up to the present
($z=0$) epoch are shown in Fig.\ \ref{diffuse} (a).  This limit is
roughly $m_s\lesssim 2\rm\ keV$. The limits for singlet dark matter in
the $L_{\nu_\alpha}\approx 0$ case decaying at $z \gtrsim 10$, shown
in Fig.\ \ref{diffuse} (b), is approximately $m_s\lesssim 20\rm\ keV$.
In this case the diffuse limit is greatly eased for the obvious reason
that there is little diffuse dark matter at recent epochs.  The true
limit from a diffuse component should lie between these extreme cases.
The actual limit depends on the epoch of signficant structure
formation, and therefore on particular structure formation models.

\section{Conclusions}
We have argued that the serendipitous coincidence betweeen the ``sweet
spot'' for x-ray photon detection technology (photon energies between
0.5 and 10 keV on modern observatories like {\it Chandra} and {\it XMM-Newton}),
and the structure consideration-preferred WDM particle rest mass
range, may afford an opportunity to detect or exclude a number of WDM
candidates.  Present observations of x-ray emission from clusters of
galaxies (e.g., Virgo) may already nearly eliminate all but a small
parameter region for the zero lepton number, non-resonant scattering
production mode for singlet ``sterile'' neutrino WDM.  We find that
the non-observance of a significant feature in deep observations of
the central region of the Virgo Cluster excludes singlet neutrino WDM
candidates with masses $m_s \gtrsim 5\rm\ keV$ in the
$L_{\nu_\alpha}\approx 0$ production mode.

However, exposures of large dark matter halos by current x-ray
observatories could yield a spectral line for $2.5 {\rm\ keV} \lesssim
m_s \lesssim 5\rm\ keV$ ($1.25 {\rm\ keV} \lesssim E_\gamma \lesssim
2.5 {\rm\ keV}$) in the $L_{\nu_\alpha}\approx 0$ case unassociated
with any atomic line.  As detailed observations of the spatially
resolved gas temperature profiles of rich clusters continues, accurate
determinations of the dark matter profile in these clusters can be
made, and existing limits and their uncertainties can be
improved. Combined with lower mass bounds, upper limits from
observation may exclude certain dark matter particle candidates.

The best strategy would be to target hot clusters or halos of spiral
galaxies, which should have few emission lines in the energy range
$1.25 {\rm\ keV} \lesssim E_\gamma \lesssim 2.5 {\rm\ keV}$.  In
addition, the inferred existence of weakly-lensing dark massive
``blobs'' \citep{Clowe2000} provides a candidate for dark matter decay
photon detection with relatively no background; however, the distance
to such dark lenses is not well known, and therefore constraints on
dark matter particle decay cannot be established from these objects.

The remaining (undetectable with ACIS on {\it Chandra})
$L_{\nu_\alpha}\approx 0$ parameter region is centered on $m_s\sim
1\rm\ keV$, the most interesting rest mass from a structure formation
standpoint.  Though this parameter space is already challenged by
potential supernova core cooling effects (see, e.g., AFP), it
would still be useful to close this window with another constraint
venue.

The nonzero lepton number ($L_{\nu_\alpha}\neq 0$) cases for singlet
neutrino WDM, corresponding to resonant production in the early
universe (\citealt{Shi:1998km}; AFP) are not as yet detectable or
constrainable with {\it Chandra}/{\it XMM}.  However, the higher
sensitivities which are possible with the Constellation X observatory
could cover much of the interesting parameter space for these singlet
neutrino models.

This is an exciting possibility.  A line feature not attributable to
an atomic line could be produced by the radiative decay of either
singlet neutrinos, heavy active neutrinos, or gravitinos.

In any case, it is remarkable and unexpected that the hard-won
technology of x-ray astronomy can in some cases probe a new regime of
particle physics, one where particle interaction strengths are some
ten orders of magnitude weaker than the Weak Interaction. Put another
way, modern x-ray obsevatories could probe epochs
of the early universe corresponding to redshifts $z\sim 10^{12}$.

\acknowledgments

We would like to thank A.\ B.\ Balantekin, J.\ Bookbinder, D.\ O.\
Caldwell, G.\ Fossati, W.\ Heindl, M.\ Patel, R.\ Rothschild, and J.\
Tomsick for useful discussions.  We would like to thank N.\ Dalal for
suggesting non-luminous massive gravitational lenses as dark matter
sources. This work was supported in part by NSF Grant PHY-9800980 at
UCSD.  K.A. would like to acknowledge support from a NASA GSRP
Fellowship. W.T. was supported in part by NASA contract NAS8-39073.

{\it Note:} It should be noted that the three-neutrino decay processes
of the singlet neutrino in Fig.\ \ref{nunufeynman} (with the rate of
equation [\ref{decayrate}]) are only possible in the presence of a
flavor-changing neutral current between neutrino flavors.  We thank
John Beacom for suggesting a clarification and emphasis of this point.

\bibliography{kev}
\bibliographystyle{apje}

\begin{deluxetable}{lrrrrrr} 
\tablecaption{Clusters of Galaxies\label{table1}}
\tablecolumns{7} 
\tablewidth{0pc} 
\tablehead{ 
\colhead{Name} & 
\colhead{$M_{\rm vir}(M_x)$\tablenotemark{a}}   & 
\colhead{$r_c$\tablenotemark{a}}    &
\colhead{$M_{\rm core}$\tablenotemark{b}} & 
\colhead{$M_{\rm DM}^{\rm fov}$\tablenotemark{c}} & 
\colhead{$\Gamma_\gamma$ Limit\tablenotemark{d}} & 
\colhead{$m_s (L_{\nu_\alpha}\approx 0)$ limit\tablenotemark{e}}
\\
\colhead{}&\colhead{$(\times 10^{14}h^{-1} M_\odot)$}&
\colhead{$(h^{-1}\rm Mpc)$}&
\colhead{$(\times 10^{14} M_\odot)$}&\colhead{$(10^{14}
M_\odot)$}&\colhead{$(\rm yr^{-1})$}&\colhead{(keV)}}
\startdata 
2A0335+096&(1.10) &0.023 &0.036\phn&1.04&$(1.1\times 10^{19})^{-1}$&4.5\\
A0085 &9.88 &0.086&0.313\phn &3.53&$(1.6\times 10^{19})^{-1}$&3.9\\
A0119 &2.50 &0.231&0.312\phn &2.23&$(1.4\times 10^{19})^{-1}$&4.1\\
A0262 &1.32 &0.032&0.029\phn &0.33&$(1.6\times 10^{19})^{-1}$&3.9\\
A0400 &2.49 &0.051&0.076\phn &0.47&$(1.0\times 10^{19})^{-1}$&4.5\\
A0426 &9.08 &0.020&0.063\phn &1.08&$(4.3\times 10^{19})^{-1}$&2.8\\
A0496 &3.20 &0.035&0.058\phn &1.40&$(1.6\times 10^{19})^{-1}$&3.9\\
A0539 &2.01 &0.082&0.094\phn &1.00&$(1.6\times 10^{19})^{-1}$&4.0\\
A1060 &1.90 &0.040&0.045\phn &0.38&$(3.7\times 10^{19})^{-1}$&2.9\\
A1656 &4.97 &0.208&0.245\phn &1.98&$(4.7\times 10^{19})^{-1}$&2.7\\
A1795 &5.86 &0.068&0.171\phn &4.20&$(1.3\times 10^{19})^{-1}$&4.2\\
A2063 &3.04 &0.067&0.110\phn &1.43&$(1.4\times 10^{19})^{-1}$&4.1\\
A2199 &5.71 &0.040&0.102\phn &1.42&$(2.0\times 10^{19})^{-1}$&3.6\\
A2256 &23.12 &0.228&1.40\phn\phn &5.26&$(1.8\times 10^{19})^{-1}$&3.8\\
A2319 &39.54 &0.135&1.24\phn\phn &4.87&$(1.9\times 10^{19})^{-1}$&3.7\\
A2634 &4.31  &0.123&0.273\phn &0.98&$(1.3\times 10^{19})^{-1}$&4.2\\
A3526 &(0.80)&0.038&0.043\phn &0.37&$(3.9\times 10^{19})^{-1}$&2.9\\
A3558 &11.54 &0.075&0.318\phn &2.22&$(1.2\times 10^{19})^{-1}$&4.3\\
A3571 &8.17  &0.086&0.241\phn &2.95&$(2.3\times 10^{19})^{-1}$&3.5\\
A4059 &(1.50) &0.075&0.164\phn &1.73&$(1.0\times 10^{19})^{-1}$&4.6\\
AWM7  &5.77   &0.062&0.148\phn &0.65&$(2.8\times 10^{19})^{-1}$&3.3\\
MKW3S &(2.00) &0.047&0.067\phn &1.83&$(1.2\times 10^{19})^{-1}$&4.3\\
MKW4  &1.15   &0.009&0.0070 &0.30&$(9.7\times 10^{18})^{-1}$&4.6\\
Virgo &2.04   &0.007&0.0081 &0.10&$(5.6\times 10^{19})^{-1}$&2.6\\
\enddata 
\tablenotetext{a}{The values of $M_{\rm
vir}$ and $r_c$ are from \citet{Horner1999}.  For clusters whose
$M_{\rm vir}$ is not well known, we used the mass determined by
the x-ray profile, $M_X$ (shown in
parentheses).}

\tablenotetext{b}{The mass within the core radius $M_{\rm core}$.}
\tablenotetext{c}{The mass within the field of view $M_{\rm DM}^{\rm
fov}$ (density integrated over the rectangular prism in the field of
view).}  
\tablenotetext{d}{The potential limit on the radiative decay
rate $\Gamma_\gamma$ of dark matter from the object with a 36 ksec
observation.}  
\tablenotetext{e}{The corresponding constraint on the
singlet neutrino mass, $m_s$, in the $L_{\nu_\alpha}\approx 0$ case
[equation\ (\ref{handy})].}
\end{deluxetable}

\begin{deluxetable}{rrrrrrr} 
\tablecolumns{7} 
\tablewidth{0pc} 
\tablecaption{Field Galaxies} 
\tablehead{ 
\colhead{Name} & \colhead{$\rho_s$}   &
\colhead{$r_s$}    
& \colhead{$M(<r_s)$} 
& \colhead{$M_{\rm DM}^{\rm fov}$\tablenotemark{a}}   
& \colhead{$\Gamma$ Limit\tablenotemark{b}} & 
\colhead{$m_s (L_{\nu_\alpha}\approx 0)$ limit\tablenotemark{c}}\\
\colhead{}&\colhead{$(\times 10^{14}\ M_\odot/\rm Mpc^3)$}&
\colhead{$(\rm kpc)$}&
\colhead{$(\times 10^{11} M_\odot)$}&\colhead{$(\times 10^{11}
M_\odot)$}&\colhead{$(\rm yr^{-1})$}&\colhead{(keV)}
}
\startdata 
NGC 3198  &1.5 &67.0&4.32 &3.62&$(2.3\times 10^{18})^{-1}$&6.0\\
NGC 4123  &1.3 &38.2&0.704 &1.85&$(2.3\times 10^{17})^{-1}$&10.3
\enddata 
\tablenotetext{a}{The mass within the field of view $M_{\rm DM}^{\rm
fov}$ (density integrated over the rectangular prism in the field of view).}
\tablenotetext{b}{The potential limit on the radiative decay rate
$\Gamma_\gamma$ of dark matter from the object with a 36 ksec
observation.}
\tablenotetext{c}{The corresponding constraint on the singlet neutrino
mass, $m_s$,in the $L_{\nu_\alpha}\approx 0$ case 
[equation\ (\ref{handy})].}
\end{deluxetable}

\end{document}